\documentclass[10pt,preprint]{aastex62}

\graphicspath{{plots/}}

\usepackage{mathrsfs, amsmath}
\usepackage{dblfloatfix}
\usepackage{csquotes}
\usepackage{dblfloatfix}
\usepackage{fixltx2e}
\usepackage{url}
\usepackage{natbib}

\usepackage{graphicx}
\usepackage{enumitem}
\usepackage{cancel}

\submitjournal{PASP}
\accepted{March 28, 2020}

\shorttitle{FAST TGV}
\shortauthors{Gerard, B. L. et al.}

\begin{document}
\title{Focal Plane Wavefront Sensing with the FAST TGV Coronagraph}

\correspondingauthor{Benjamin L. Gerard}
\email{bgerard@uvic.ca}

\author[0000-0003-3978-9195]{Benjamin L. Gerard}
\affil{University of Victoria, Department of Physics and Astronomy, 3800 Finnerty Rd, Victoria, V8P 5C2, Canada}
\affiliation{National Research Council of Canada, Astronomy \& Astrophysics Program, 5071 West Saanich Rd, Victoria, V9E 2E7, Canada}

\author[0000-0002-4164-4182]{Christian Marois}
\affiliation{National Research Council of Canada, Astronomy \& Astrophysics Program, 5071 West Saanich Rd, Victoria, V9E 2E7, Canada}
\affiliation{University of Victoria, Department of Physics and Astronomy, 3800 Finnerty Rd, Victoria, V8P 5C2, Canada}
\begin{abstract}
The continual push to directly image exoplanets at lower masses and closer separations orbiting around bright stars remains limited by both quasi-static and residual adaptive optics (AO) aberration. In previous papers we have proposed a modification of the self-coherent camera (SCC) design to address both of these limitations, called the Fast Atmospheric SCC Technique (FAST). In this paper we introduce an additional modification to the FAST focal plane mask design, including the existing Tip/tilt and Gaussian components and adding a charge four Vortex (TGV) component. In addition to boosting SCC fringe signal-to-noise ratio (S/N) as in our previous design, we show that the FAST TGV mask is also optimized to reach high contrast at separations closer to the star. In this paper we use numerical simulations to consider the performance improvement on correcting quasi-static aberration using this new mask compared to the previously proposed Tip/tilt+Gaussian mask. Using active deformable mirror control to generate a calibrated half dark hole improves contrast by a factor of about 200 at 2 - 5 $\lambda/D$ and up to a factor of 10 at 5 - 20 $\lambda/D$. The new methodology presented in this paper, now simultaneously considering both contrast and fringe S/N, opens the door to a new ideology of coronagraph design, where the coronagraph is now considered in duality as both a diffraction attenuator and a wavefront sensor.
\end{abstract}
\section{Introduction}
\label{sec: intro}
Direct imaging of exoplanets is a key science goal of both current and future ground- and space-based observatories. Continual advancements in both hardware and software have allowed newer instruments to be more sensitive than their predecessors to exoplanets lower in mass and closer in orbital separations, arising from reaching deeper contrasts and smaller inner working angles (IWAs), respectively. However, both temporal and chromatic aberrations in high contrast imaging instruments are currently limiting the magnitude of these improvements \citep[e.g.,][]{cor1,cor2,cor3,me_scexao}.

\cite{fast} and \cite{fast_spie} (hereafter G1 and G2, respectively) proposed a solution to remove these temporal limitations that arise from both quasi-static and dynamic aberrations, called the Fast Atmospheric Self-Coherent Camera (SCC) Technique (FAST). The SCC was invented by \cite{scc_orig} and then modified by \cite{scc_lyot}, using a multi-aperture design in the Lyot stop plane (i.e., a common-path interferometer) to spatially modulate speckles and diffraction in the coronagraphic image. Isolating the recorded fringes in the Fourier domain then allows for a measurement of the complex focal plane electric field of the star without bias from any planet light; light from an off-axis planet beyond the coronagraph IWA is not fringed in the coronagraphic image because only starlight is diffracted into the off-axis Lyot stop pinhole, while the planet light is transmitted only through the central pupil. If the stellar speckles are generated from dynamic aberrations, fringes must be detected on a speckle with only a few recorded photons, or at a relatively high fringe signal-to-noise ratio (S/N), before the electric field evolves into another uncorrelated realization. The core component of FAST is a new focal plane mask (FPM) that is designed to increase the fringe S/N of the SCC to allow for such a detection. This fringe S/N boost enables wavefront sensing and subtraction of both quasi-static and residual adaptive optics (AO) speckles by post-processing (G1) and/or deformable mirror (DM) control (G2). The FPM proposed in G1, however, was not optimized for contrast, following a simple Lyot coronagraph design that is limited by bright diffraction rings in the coronagraphic image. Although speckles ``pinned'' by diffraction rings \citep{perrin_psf} can be well-subtracted by post-processing (G1), this poses significant difficulties in DM control and adds additional photon noise to the coronagraphic image that cannot be removed by post-processing (G2).

In this paper we propose a modification of the original FAST FPM that is better optimized for both fringe S/N and attenuation of diffraction: the Tip/tilt, Gaussian, Vortex (TGV) FPM. In \S\ref{sec: tgv} we introduce the design of the TGV mask and present a new framework behind FAST coronagraph design, now treating the FPM as both a diffraction attenuator \textit{and} a wavefront sensor (WFS). Then in \S\ref{sec: dm_control} we show the relative advantages of dark hole generation using the TGV mask compared to previous work, and finally in \S\ref{sec: conclusions} we conclude and discuss the future outlook from this paper. Unless otherwise noted, the numerical setup for our simulations is the same as described in G1, including the parameters listed in Table \ref{tab: setup} below. Further setup and parameter definitions are provided in appendices \ref{sec: scc} - \ref{sec: fringe_snr}.
\begin{table}[!h]
\begin{tabular*}{\textwidth}{@{\extracolsep{\fill}} llll}
variable & value & unit & explanation \\
\hline
$\lambda$ & 1.65 & $\mu$m & wavelength of light (monochromatic, no photon noise) \\
N$_\text{act}$ & 32 & dimensionless & number of DM actuators across the entrance pupil \\
$\xi_0$ & 1.54 & entrance pupil radii & distance between the center of the Lyot pupil and SCC pinhole\\
beam ratio & 5.4 & pixels & number of pixels per $\lambda/D$ resolution element \\
D$_\text{Lyot}$ & 0.95 & dimensionless & Lyot stop diameter in fraction of the entrance pupil diameter \\
D$_\text{pin}$ & 1/18.5 & dimensionless & SCC pinhole diameter in fraction of the entrance pupil diameter \\
pl$_\phi$ & -1.5 & dimensionless & power law assumed for quasi-static phase aberration \\
${\text{pl}_{\phi_a}}$ & -2 & dimensionless & power law assumed for residual AO phase aberration \\
pl$_a$ & -2 & dimensionless & power law assumed for quasi-static amplitude aberration \\
\end{tabular*}
\caption{SCC variables used in this paper.}
\label{tab: setup}
\end{table}
\section{The TGV FPM}
\label{sec: tgv}
Building on the design proposed in G1, the main goal of our proposed FPM modification in this paper is improve the diffraction-limited contrast in the coronagraphic image while still maintaining a sufficient fringe S/N. The latter fringe S/N requirement remains essential in order to operate the FAST technique on millisecond timescales for bright stars. Although the raw contrast, dominated by unpinned quasi-static and/or atmospheric speckles, may not show an improvement over the previous design, we ultimately want to reach a deeper contrast in the subtracted image by post-processing and/or DM control. With this in mind, the TGV FPM is a focal plane phase mask with three components, each described below:
\begin{itemize}[wide=0.5em,leftmargin=2em]
\item[Tip/tilt:] generate a spatially filtered, off-axis pupil in the Lyot plane (the off-axis pupil is limited to Fourier modes less than $2e$ cycles/pupil, where $e$ represents the radius of the central Tip/tilt+Gaussian region in $\lambda/D$),
\item[Gaussian:] concentrate, in the Lyot plane, intensity on the off-axis pupil generated by the Tip/tilt component, and
\item[Vortex:] redistribute diffracted star light from inside to outside the central Lyot pupil using a vortex phase ramp \citep{vortex}.
\end{itemize}
\begin{figure}[!h]
\centering
\includegraphics[width=1.0\textwidth]{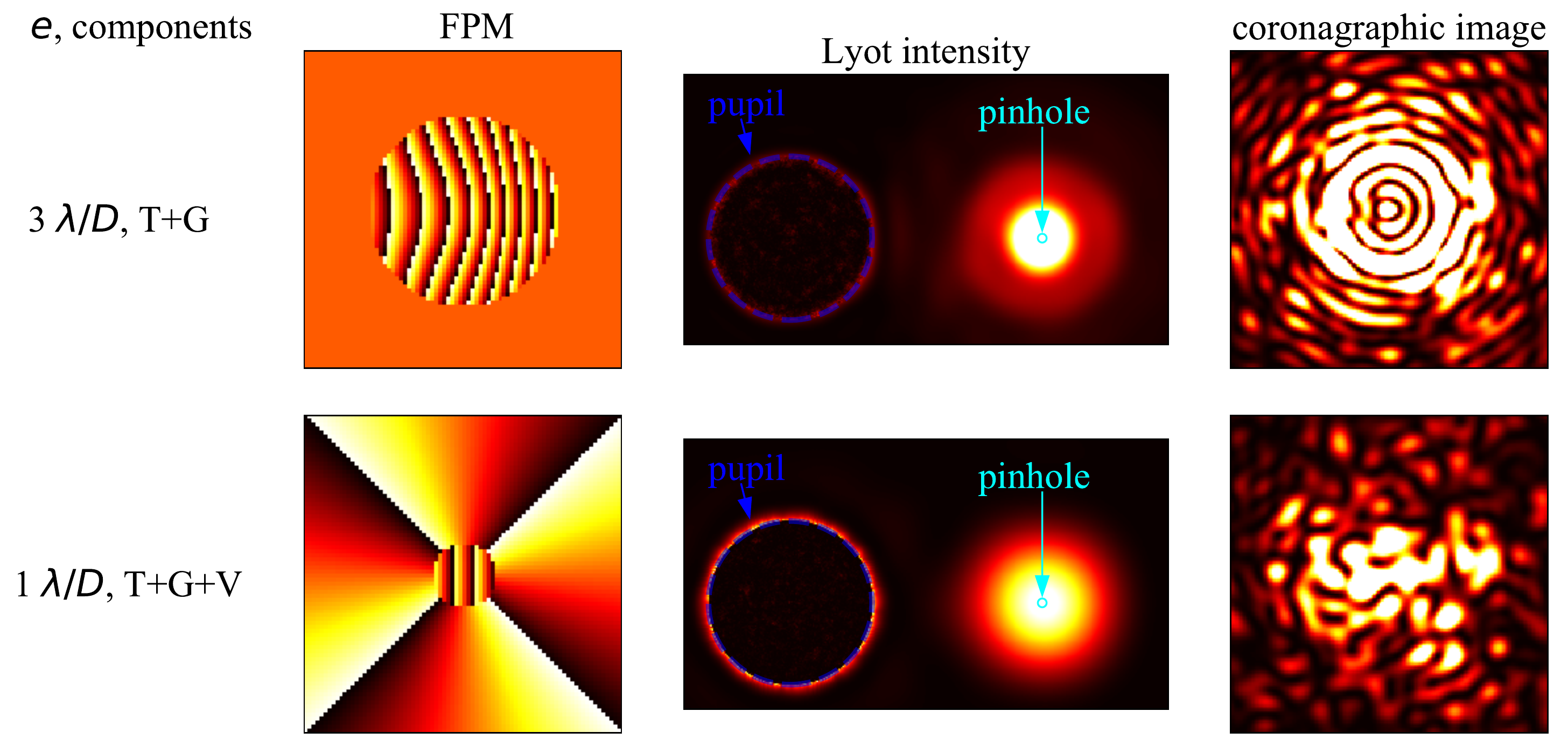}
\caption{A comparison of the old (TG, top row) and new (TGV, bottom row) coronagraphic phase masks, showing the phase-wrapped optical path difference (left ``FPM'' column), corresponding Lyot plane intensity before a Lyot stop is applied (middle ``Lyot intensity'' column), and corresponding coronagraphic image after a Lyot stop is applied with no SCC pinhole (right column). The pinhole and entrance pupil apertures in the Lyot plane are illustrated in cyan and blue, respectively. Each column is on the same contrast scale, and the same single aberrated wavefront realization, applied in the entrance pupil, is used to generate both Lyot intensity and coronagraphic images above. }
\label{fig: tgv_imas}
\end{figure}

Figure \ref{fig: tgv_imas} conceptually illustrates the differences between our old TG mask and our new TGV mask. First, considering only the intensity in the off-axis Lyot pupil, a larger $e$ value concentrates more light into the SCC pinhole. This effect occurs from the optical relationship between the FPM and pupil plane diameters; for $e=1\; \lambda/D$, light concentration of the off-axis Lyot pupil, enabled by the $G$ term, can only decrease the full width at half maximum (FWHM) by a factor of 2 (i.e., the diffraction limit), whereas for $e=3\; \lambda/D$, the downstream pupil FWHM can instead shrink by a factor of 6. In addition to this ``compresibility'' diffraction limit, a larger $e$ value collects more light from the on-axis star that is relayed into the off-axis pupil. Second, considering only the intensity distribution around the central Lyot pupil, more diffracted light is sent outside of the pupil by the TGV mask than the TG mask, illustrated by the bright ring around the edge of the pupil generated by the TGV mask. This effect is then seen in the coronagraphic image generated using the TGV mask, producing speckles that are no longer pinned to bright diffraction rings. The attenuation of pinned speckles by the TGV mask occurs because more diffracted light is redistributed outside of the central pupil and blocked by the Lyot stop, while dynamic and quasi-static aberrations from the telescope and instrument, respectively, are still transmitted through the central pupil.

The mathematical prescription for the TGV mask, $\phi_\text{TGV}$, is defined as follows:
\begin{align}
T &\equiv3.16(\xi_0)(x\;\text{cos}\theta_0+y\; \text{sin}\theta_0),\; \forall\ r<e\nonumber \\ 
G &\equiv g\; e^{-\frac{1}{2}\left(\frac{r}{\sigma}\right)^2},\; \forall\; r<e \nonumber \\ 
\label{eq: tgv} V &\equiv l_p \;\theta,\; \forall\; r>e, \\
\phi_\text{TGV} &= T+G+V\; [rad] \nonumber
\end{align}
where $x$ and $y$ are a linear ramp in units of $\lambda/D$ along each respective axis with the zero point corresponding to the optical axis, $r \equiv \sqrt{x^2+y^2},\; \theta \equiv \text{tan}^{-1}\left(\frac{y}{x}\right)$, $\xi_0$ is the distance between the center of the Lyot pupil and the center of the pinhole in units of pupil radii, and $\theta_0$ is the position angle of the tip/tilt direction applied to the $T$ component (counter-clockwise from the +$x$ direction) and optically matched to the corresponding position angle between the centers of the pupil and SCC pinhole in the Lyot plane. Conceptually, the effect of the $G$ term is to emulate a speckle or point spread function (PSF) core whose resolution limit is larger than $\lambda/D$, corresponding to a spatially filtered pupil in the Lyot plane whose FWHM is smaller than the re-maped entrance pupil. We found that a symmetric 2D Gaussian function is effective at concentrating light in the Lyot plane, although others similar functions could be examined in future global optimizations (see \S\ref{sec: conclusions}). Values for $g,$ $\sigma$, and $l_p$ represent the Gaussian amplitude (in radians), Gaussian width (in $\lambda/D$), and the integer-valued topological charge of $V$ (dimensionless; \citealt{vortex}), respectively, and are all free parameters to be optimized. For a chosen value of $e$ we performed a grid search optimization of $g$ and $\sigma$ using the Lyot plane intensity distribution; integrated intensity over the pinhole divided by integrated intensity over the pupil was computed for every grid value, and the chosen optimal values were set to optimize this integrated Lyot fringe ratio metric. We did not run an additional grid search for $e$ and $l_p$. Using this procedure we found an optimal TGV prescription of $e=1\; \lambda/D$, $g=6\; rad$, $\sigma=5\; \lambda/D$, and $l_p=4$. In G1 and G2, the TG mask was defined with $e=3\; \lambda/D$, $g=11.5\; rad$, $\sigma=2.0\; \lambda/D$, and $l_p=0$.
\begin{figure}[!ht]
\centering
\includegraphics[width=1.0\textwidth]{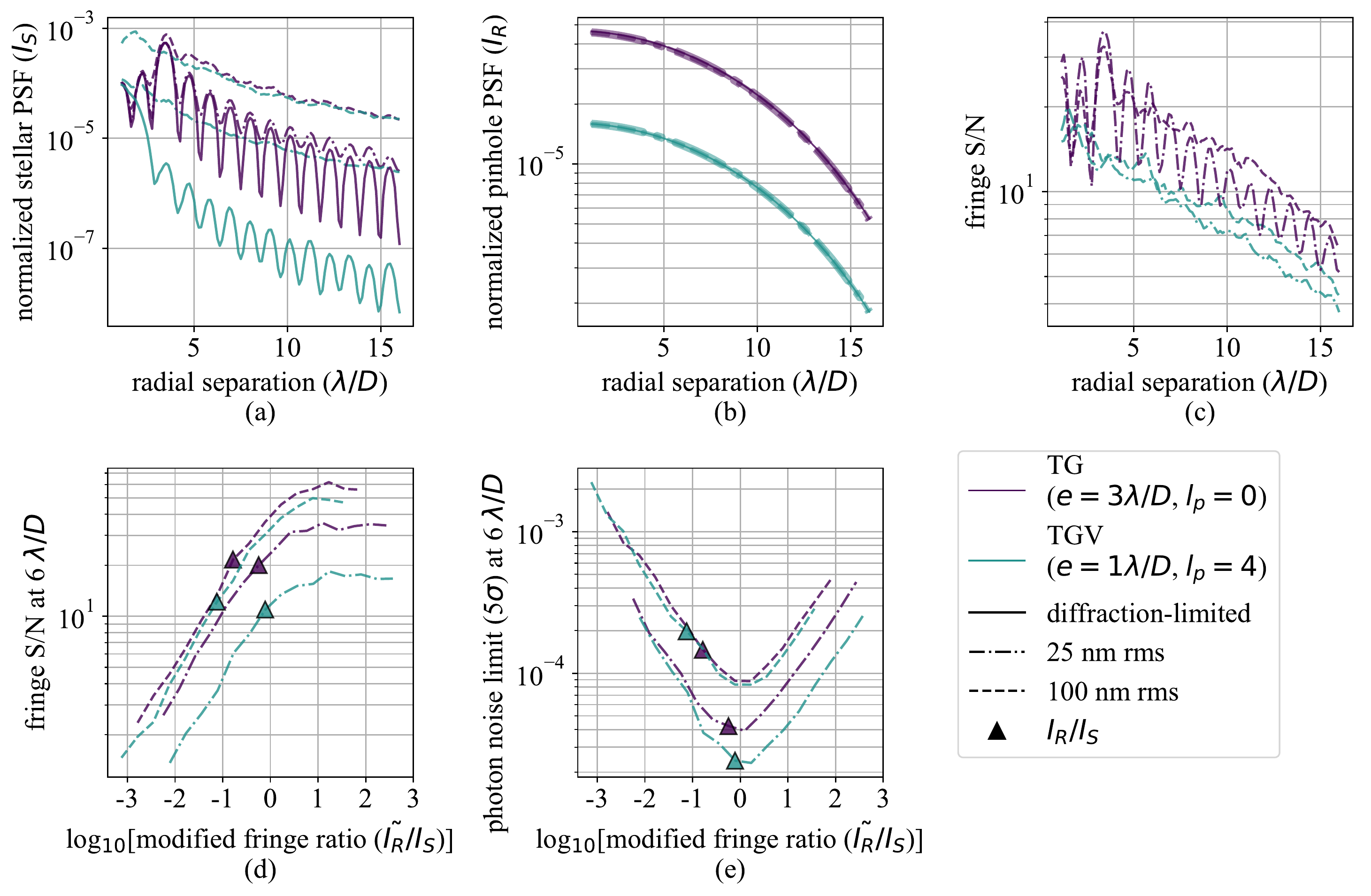}
\caption{An illustration of the tradeoffs between coronagraphic image intensity (a), pinhole PSF intensity (b), fringe S/N (c and d), and photon noise-limited contrast (e), utilizing the definitions in Appendices \ref{sec: scc} - \ref{sec: fringe_snr}. The purple and teal curves represent the old TG and new TGV masks, respectively. The solid, dotted, and dashed-dotted lines represent performance with no wavefront error (i.e., diffraction-limited in both phase and amplitude), on-sky conditions for a 10 millisecond exposure with both AO residuals and quasi-static wavefront errors (i.e., 100 nm rms phase aberration and 1 \% rms intensity aberration), and on-sky conditions for a 10 millisecond exposure with a perfect AO correction but remaining uncorrected non-common path errors (i.e., 25 nm rms phase aberration, 1 \% rms intensity aberration), respectively. For the curves simulating on-sky conditions, each line shown is the median contrast or fringe S/N curve, each first individually determined from 10 different uncorrelated entrance pupil wavefront realizations. See Appendix \ref{sec: fringe_snr} for a definition of ``fringe ratio'' and ``modified fringe ratio.'' The fringe ratio values in panels d and e are indicated for each curve by a triangle symbol of the corresponding color.}
\label{fig: vortex_wfe}
\end{figure}

\begin{figure}[!ht]
\centering
\includegraphics[width=0.5\textwidth]{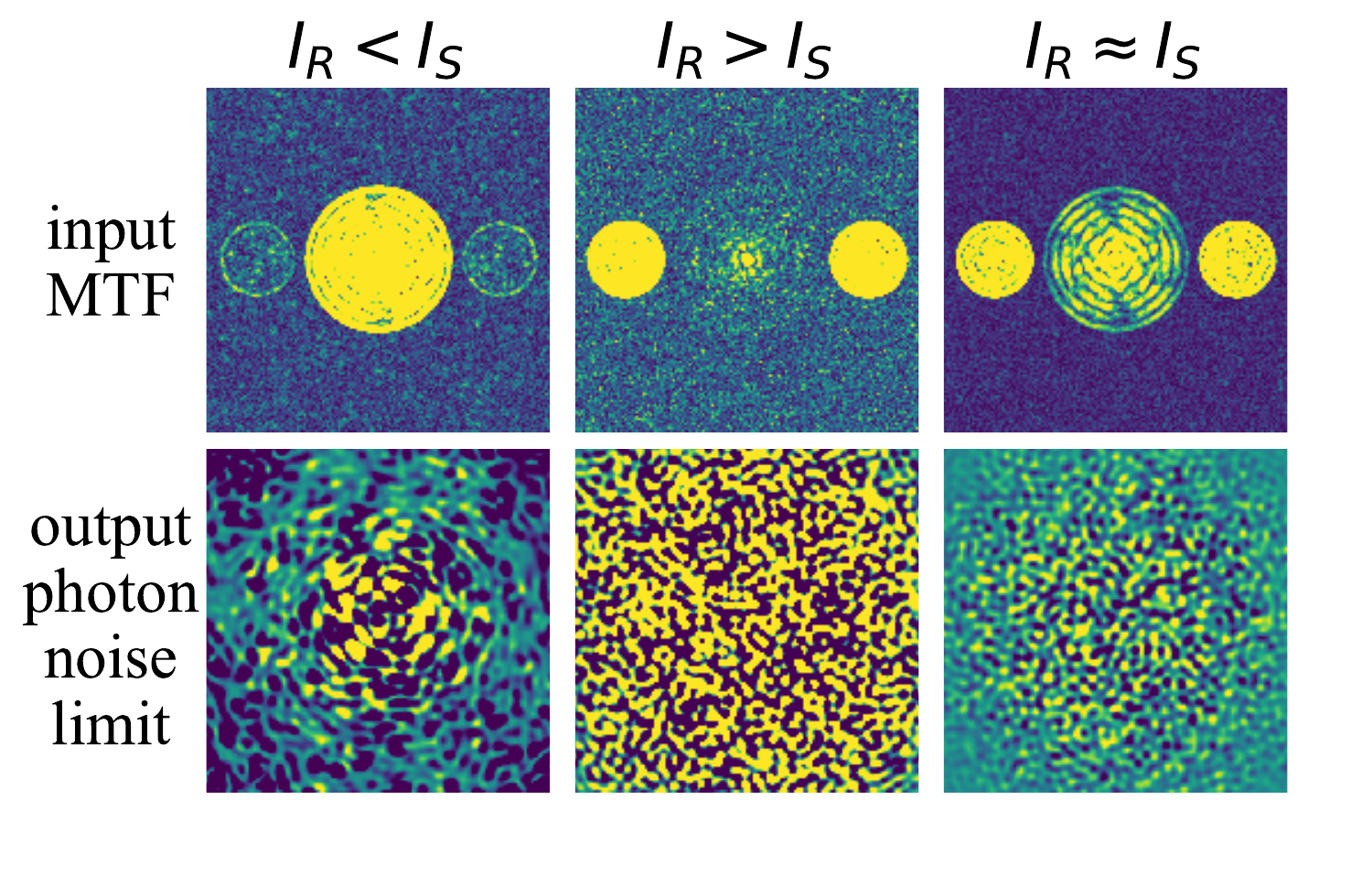}
\caption{An illustration of three different modified fringe ratio cases, building off of Fig. \ref{fig: vortex_wfe} d and e, showing for each case the input image modulation transfer function (MTF) and the output photon noise limit (images in this bottom row are all shown on the same linear contrast scale), both as defined in Appendix \ref{sec: fringe_snr}. SCC images are generated by simulating a 10 ms on-sky exposure (i.e., with photon noise and the setup described in Appendix \ref{sec: contrast}), using the TGV FPM and adjusting the Lyot plane pinhole intensity as described in Appendix \ref{sec: fringe_snr}.}
\label{fig: phlim_illustration}
\end{figure}

Utilizing the setup and definitions in appendices \ref{sec: scc} - \ref{sec: fringe_snr}, Fig. \ref{fig: vortex_wfe} provides a more robust illustration of the tradeoffs between contrast and fringe S/N, and illustrates a new approach to coronagraph design, simultaneously considering diffraction attenuation, WFS sensitivity, and photon noise-limited contrast, shown in Fig. \ref{fig: vortex_wfe} a, c/d, and e, respectively. The main conclusion from Fig. \ref{fig: vortex_wfe}, illustrated in panel e, is that photon noise-limited contrast is optimized at a fringe ratio of $I_R/I_S\approx1$. This concept is illustrated further in Fig. \ref{fig: phlim_illustration}.
Summarizing the results from Figures \ref{fig: vortex_wfe} d and e and Fig. \ref{fig: phlim_illustration}, the two less optimal regimes of photon noise-limited contrast are:
\begin{itemize}[leftmargin=0.75in]
\item[$I_S>I_R$:] An insufficient amount of light is sent through the Lyot stop pinhole, such that excess photon noise from $I_S$ both decreases the fringe S/N and increases the photon noise-limited contrast (left column of Fig. \ref{fig: phlim_illustration}). This scenario is the most common for typical coronagraph designs, where fringe ratio is not optimized in the design procedure.
\item[$I_S<I_R$:] Too much light is being sent through the Lyot stop pinhole, such that excess photon noise from $I_R$ degrades the photon noise-limited contrast (middle column of Fig. \ref{fig: phlim_illustration}; any exoplanet which could be detected in the central lobe of the MTF in the upper left or right panels is now buried in the photon noise generated from $I_R$), although Fig. \ref{fig: vortex_wfe} d clearly shows that this effect does not decrease fringe S/N, which instead asymptotes as $I_R$ continually increases in this regime. However, if new coronagraph designs can enable this regime, note that this case can be mitigated by modifying the size \citep{mazoyer}, transmission, and/or complex electric field imparted through the Lyot stop pinhole, adjusting $I_R$ to best match the amplitude of  $I_S$.
\end{itemize}
\noindent Thus, incorporating the tradeoffs with contrast (see below, which will determine the WFS sensitivity to non-linearities; \citealt{coron}), $I_R/I_S=1$ should be adopted as a coronagraph design parameter to optimize WFS sensitivity to photon noise \citep{guyon_ao} without degrading the achievable photon noise-limited contrast.

Additional conclusions from Fig. \ref{fig: vortex_wfe} are similar to those from Fig. \ref{fig: tgv_imas}, showing that 
\begin{enumerate}
\item for open-loop (i.e., FAST loop open, AO loop closed) fast on-sky exposures that ``freeze'' both the atmospheric residuals and quasi-static aberration (i.e., curves labeled ``100 nm rms''), the fringe S/N is generally higher for the old TG design than for the new TGV design due to the higher level of $I_R$ in panel b while contrast is the same for both (panel a), but 
\item the diffraction-limited contrast is orders of magnitude better for the new TGV vs. old TG mask design. 
\end{enumerate}
As a result, raw contrasts for the ``25 nm rms'' case are lower for the TGV than the TG design, and accordingly the TGV mask reaches a deeper photon noise-limited contrast for this case. For the 100 nm rms case, photon noise-limited contrast is instead lower for the TG mask, due to the aforementioned higher fringe S/N but equal contrast levels compared to the TGV mask. This suggests that, for on-sky DM control of un-pinned speckles (i.e., minimizing entrance pupil wavefront error), there is a crossing point once the FAST loop is closed where the achievable photon noise-limited contrast of the TGV mask surpasses the values of the TG mask. Thus, although fringes for the TGV design would be detected at a relatively lower S/N in open loop millisecond frames, deeper contrasts are expected if the FAST loop can close (see \S\ref{sec: conclusions} for further discussion).
\section{Quasi-Static DM control}
\label{sec: dm_control}
\begin{figure}[!h]
\centering
	\begin{minipage}[b]{0.5\textwidth}
		\begin{center}
		\includegraphics[width=1.0\textwidth]{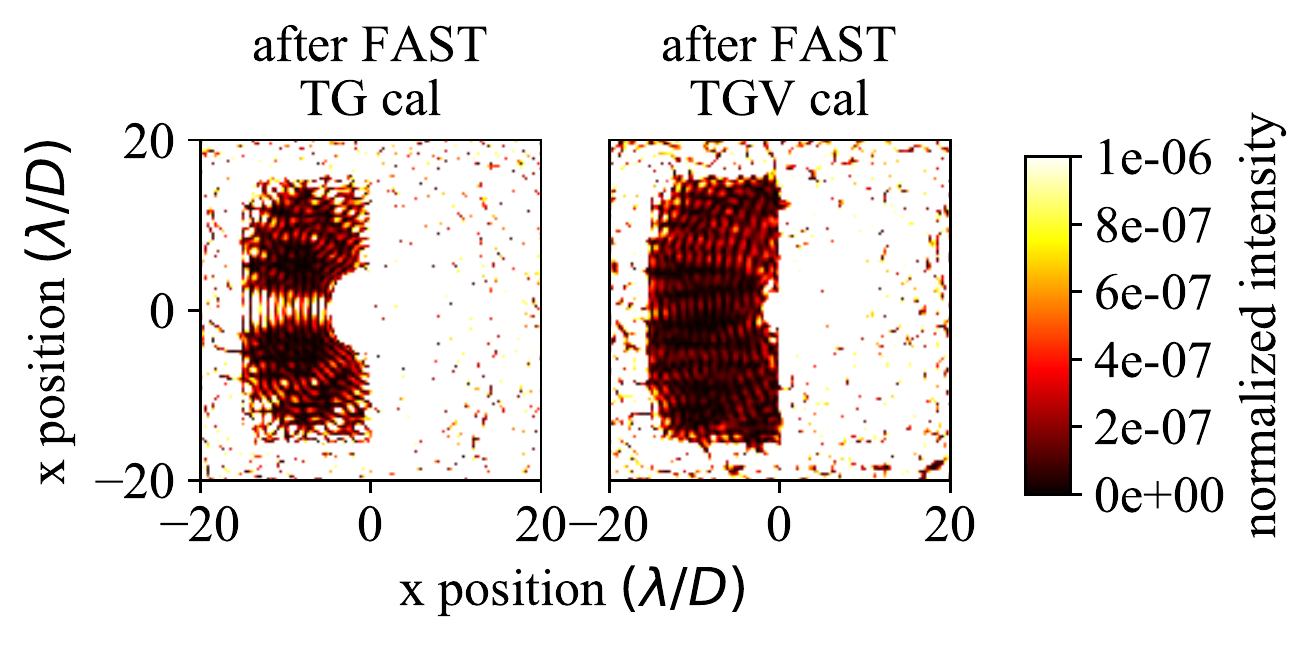}
		(a)
		\label{fig: a}
		\end{center}
	\end{minipage}
	\begin{minipage}[b]{0.4\textwidth}
		\begin{center}
		\includegraphics[width=1.0\textwidth]{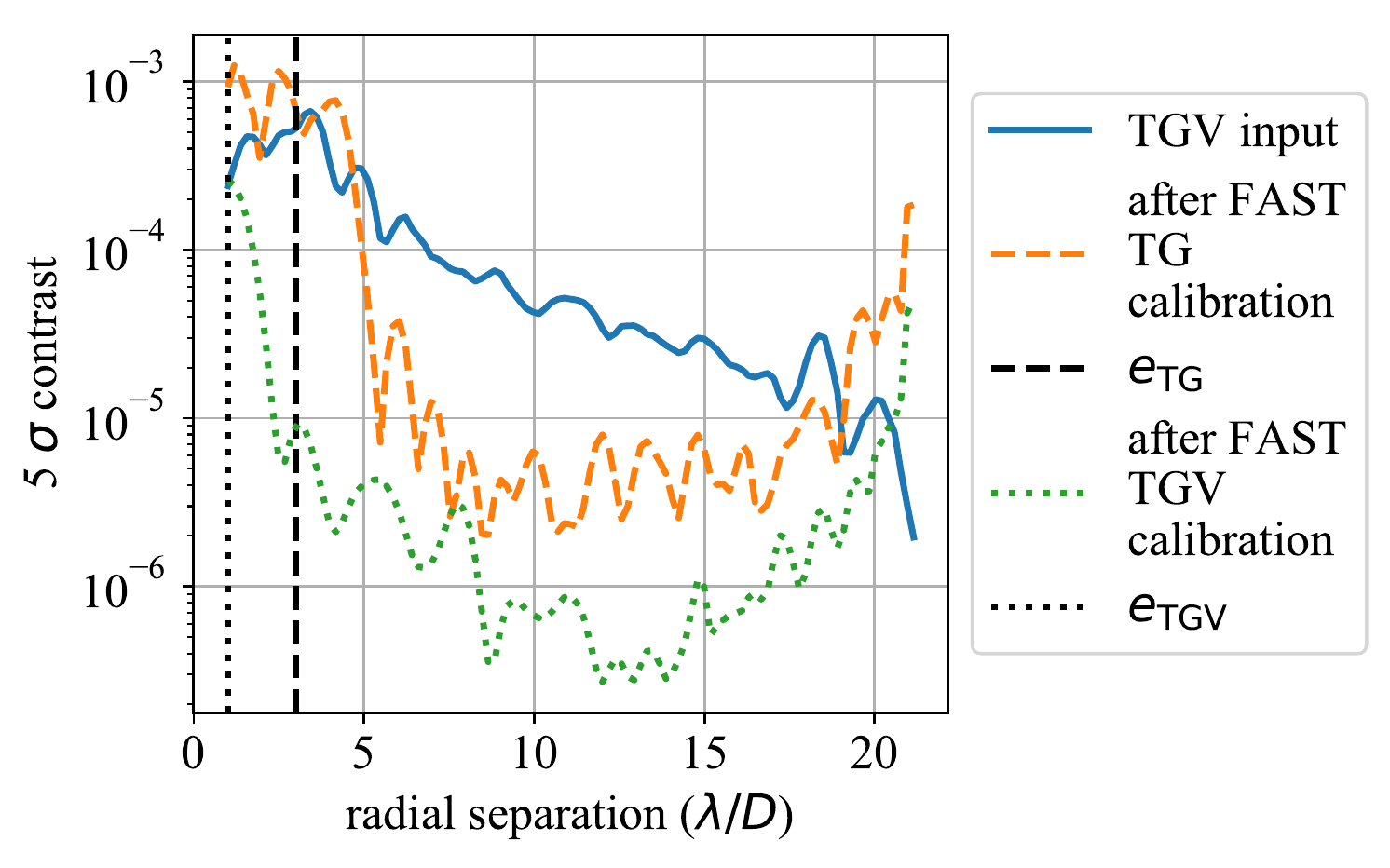}
		(b)
		\label{fig: b}
		\end{center}
	\end{minipage}	
\caption{(a) Calibrated half dark hole $I_S$ images (i.e., Lyot stop pinhole closed) using the old TG and new TGV mask design (left and right panels, respectively). (b) Contrast curves for the two images in (a), calculated on pixels only within the half dark hole, as well as for the TG input image before calibration. The same static wavefront realization (i.e., 25 nm rms phase and 1 \% rms intensity aberration) is input into both the TG and TGV calibration procedures. }
\label{fig: dh}
\end{figure}

Here we examine the performance of the TGV mask in generating a half dark hole (DH) via DM control of quasi-static speckles. As a reminder, the main error terms that FAST addresses in enabling deeper detections are quasi-static and residual AO speckles. Note that the SCC command matrix relies on a linear assumption to transform SCC images into DM commands in a single least-squares-based matrix multiplication \citep{baudoz_psfsubt}. Although here we are only considering correction of quasi-static aberration, we have not considered other iterative non-linear DM control algorithms that are more optimized for diffraction attenuation \citep[e.g.,][]{speckle_nulling, efc, stroke_min} as the ultimate goal of this approach will to be to run the same correction on-sky on noisy millisecond exposures. A detailed description of the SCC DM calibration procedure can be found in G2 and references therein. In Fig. \ref{fig: dh} we compare the results of this procedure for the TG and TGV masks using a single static wavefront realization as the input with an input wavefront error including 25 nm rms of phase aberration and 1\% intensity rms of amplitude aberration. The calibration results are shown after three iterations using an integrator controller and a unity gain. The control algorithm linking the Fourier modes recorded in the SCC image and DM commands relies on a Taylor expansion of the wavefront assuming small phase and amplitude defects (\citealt{baudoz_psfsubt}; i.e., linearizing $a\;e^{i\;\phi}$ to $a(1+i\; \phi)$, where $a$ and $\phi$ represent the spatial distributions of amplitude and phase, respectively, in the complex electric field of the entrance pupil plane). Multiple iterations are therefore still required, even with a unity gain, to address non-linearities between the two planes.

A comparison of the calibrated DH generated from these two coronagraphs yields a few important results:
\begin{enumerate}
\item the DH contrast with TGV mask is about 200 times deeper than the TG mask at 2 - 5 $\lambda/D$, enabled by the smaller value of $e$ and deeper diffraction-limited contrast at these separations, and
\item up to 10 times deeper at 5 - 20 $\lambda/D$.
\end{enumerate}
Note that contrast curves in Fig. \ref{fig: dh} b are shown within the value of $e$; although planet throughput at these separations is $\sim$0, the curves are still shown to illustrate the independent concept of diffraction suppression between the two coronagraphs. As discussed in G2, even though $e=3\; \lambda/D$ for the TG mask, we found that we had to use an algorithmic mask to block the central 5 $\lambda/D$ in radius because of bright diffraction in the coronagraphic image that otherwise biased the least-squares algorithm. If we instead used an algorithmic mask down to, e.g., 3 $\lambda/D$, we could not reach the same contrasts from 5 - 20 $\lambda/D$ as in Fig. \ref{fig: dh} b. We also found the same effect for the TGV mask, requiring an algorithmic DH mask with an innermost radial separation IWA of 2 $\lambda/D$ instead of the TGV $e$ value of 1. Regardless of this limitation, the TGV mask clearly provides a gain in achievable DH contrast over the TG mask. Again, more detailed end-to-end simulations, incorporating closed-loop FAST performance, will be presented in a future paper.
\section{Conclusions and Discussion}
\label{sec: conclusions}
In this paper we have introduced the Tip/tilt + Gaussian + Vortex (TGV) focal plane mask for use with the Fast Atmospheric Self-coherent camera Technique (FAST; G1 and G2). In doing so, we have also introduced a new methodology towards coronagraph design, where contrast and fringe S/N are considered simultaneously. We have shown that the TGV mask has a number of advantages over the TG mask previously proposed in G1, including
\begin{enumerate}
\item better balance of diffraction-limited contrast, fringe S/N, and photon noise-limited contrast (\S\ref{sec: tgv}) and
\item almost 6 magnitudes deeper contrast at 2 - 5 $\lambda/D$ by DM control of quasi-static aberration (\S\ref{sec: dm_control}).
\end{enumerate}

Although we did not yet specifically address the achievable contrasts for closed-loop FAST operation using millisecond-timescale on-sky images, the framework already presented in this paper provides a promising outlook for expected on-sky performance. Even though in Fig. \ref{fig: vortex_wfe} we showed that TGV fringe S/N is above 10 at separations less than about 6 $\lambda/D$ for a 10 ms exposure with 100 nm rms wavefront error, as soon as the FAST DM control loop is closed the wavefront error should decrease to a much lower value, thereby boosting the fringe ratio to a more optimal value and improving the achievable photon noise-limited contrast. If the fringe ratio is boosted to greater than one (thereby degrading the photon noise limit), the pinhole size and/or throughput can be adjusted to set $I_R=I_S$. Thus, the main potential limitation will be whether or not the FAST DM control loop can close at the lower frame rate needed to detect fringes in the raw images; this will be investigated in detail in a forthcoming paper.

We have also illustrated that coronagraph mask design optimization is clearly a crucial step in optimizing the achievable contrast of FAST post-processing and DM control. Many factors need to be considered in the design process, such as optimizing the tradeoffs between diffraction-limited contrast, on-sky millisecond-timescale fringe S/N, and sensitivity to low-order aberrations. The initial study in this paper is meant to provide the conceptual framework for optimization of a more instrument-ready FAST coronagraph design. Future work on this topic will consider additional factors that would influence realistic coronagraph design, such as sensitivity to secondary obscuration and supports, chromaticity, and a full Monte Carlo analysis of TGV free parameters. Additionally, such optimizations will need to consider AO performance, a new approach to coronagraph design; Figure \ref{fig: vortex_wfe} has illustrated that the requirements for fringe S/N (which will trace the WFS sensitivity photon noise propagation; \citealt{guyon_ao}) and contrast (which will trace the WFS non-linearities; \citealt{coron}) are inherently tied to FAST coronagraph design. Such future FAST optimizations will therefore also need to consider the relative tradeoffs of these factors in a focal plane wavefront control AO error budget analysis, which we will explore in detail in a future paper.
\section*{Acknowledgements}
We gratefully acknowledge research support of the Natural Sciences and Engineering Council of Canada through the Postgraduate Scholarships-Doctoral award, Discovery Grant, and Technologies for Exo-Planetary Science Collaborative Research and Training Experience programs. We thank Rapha\"el Galicher for comments, suggestions, and discussions that have significantly improved this manuscript. We also thank Pierre Baudoz, Johan Mazoyer, Garima Singh, and J.-P. V{\'e}ran for helpful discussions and suggestions. The authors thank the anonymous referee for his or her comments and suggestions that have significantly improved this manuscript.
\appendix
\section{SCC Notation}
\label{sec: scc}
Using the same notation from \cite{scc_orig} and subsequent papers, $I_R$ is the intensity in the coronagraphic image plane with the light transmitted through the off-axis Lyot stop pinhole while the central Lyot stop pupil is blocked (i.e., the ``pinhole PSF''), and $I_S$ is the coronagraphic image where light is transmitted through the central Lyot stop pupil and the off-axis pinhole is blocked, and the amplitude of the SCC fringe term recorded in the coronagraphic image is $2\sqrt{I_S I_R}$ \citep{scc_orig}. The full SCC coronagraphic image can be represented by 
\begin{equation}
I=I_P+I_S+I_R+2\sqrt{I_S\; I_R}M,
\label{eq: scc_im}
\end{equation}
where $I_P$ is the off-axis exoplanet PSF and $M$ represents the spatial distribution of the fringe term, varying between $\pm1$, and is a function of wavelength, Lyot stop pinhole separation, and the differential complex electric field between the Lyot stop pinhole and pupil. There is no $I_P$ term in the fringe term because fringes are not recorded on the exoplanet, as its light does not go through the Lyot stop pinhole. The fringe term can be algorithmically isolated and filtered in the complex Fourier plane of the image, generating an image plane phase and intensity distribution known in the literature as  $I_-$, whose monochromatic amplitude is half of the fringe term amplitude: $|I_-|=\sqrt{I_R\; I_S}$.
\section{Contrast and Throughput Definitions}
\label{sec: contrast}
In \S\ref{sec: dm_control}, a contrast curve is produced by computing five times the standard deviation in an azimuthal annulus (of width 3 pixels) at a given separation in the coronagraphic image (i.e., $I_S$), normalized by the peak value of the same wavefront(s) if the FPM is removed. In \S\ref{sec: tgv}, normalized intensity curves are computed with the median intensity instead of standard deviation, otherwise using the same normalization and azimuthal bins. We chose to show intensity in \S\ref{sec: tgv} instead of standard deviation to allow comparison with the diffraction-limited case in Fig. \ref{fig: vortex_wfe}; in this case, because of azimuthal symmetry computing standard deviation is less physically representative of contrast. Also note that we compute contrast and intensity curves on only $I_S$, as opposed to the full SCC PSF, to isolate the effects of diffraction and speckle suppression vs. fringe S/N; although $I_R$ \text{will} ultimately play a role in contrast, this effect can in principle be fully attenuated in post-processing by measuring the ``live'' pinhole PSF (G1), and so we do not discuss this impact here.

As in G1 and G2 (see references therein), in \S\ref{sec: tgv} and appendix \ref{sec: fringe_snr} below we calculate the number of photons collected at the telescope entrance pupil by simulating a $m_H=0$, 1\% bandpass, 8 m telescope diameter, and 10 millisecond exposure time. The 1\% bandpass is used only for photon counting purposes in an otherwise monochromatic Fraunhofer simulation; FAST broadband operation will be explored in a future paper. Additional throughput values assumed for atmospheric transmission, transmission through telescope and instrument, and detector quantum efficiency are 90\%, 20\%, and 80\%, respectively.
\section{Fringe S/N and Photon Noise}
\label{sec: fringe_snr}
\begin{figure}[!h]
\begin{minipage}[b]{0.52\textwidth}
\begin{center}
\includegraphics[width=1.0\textwidth]{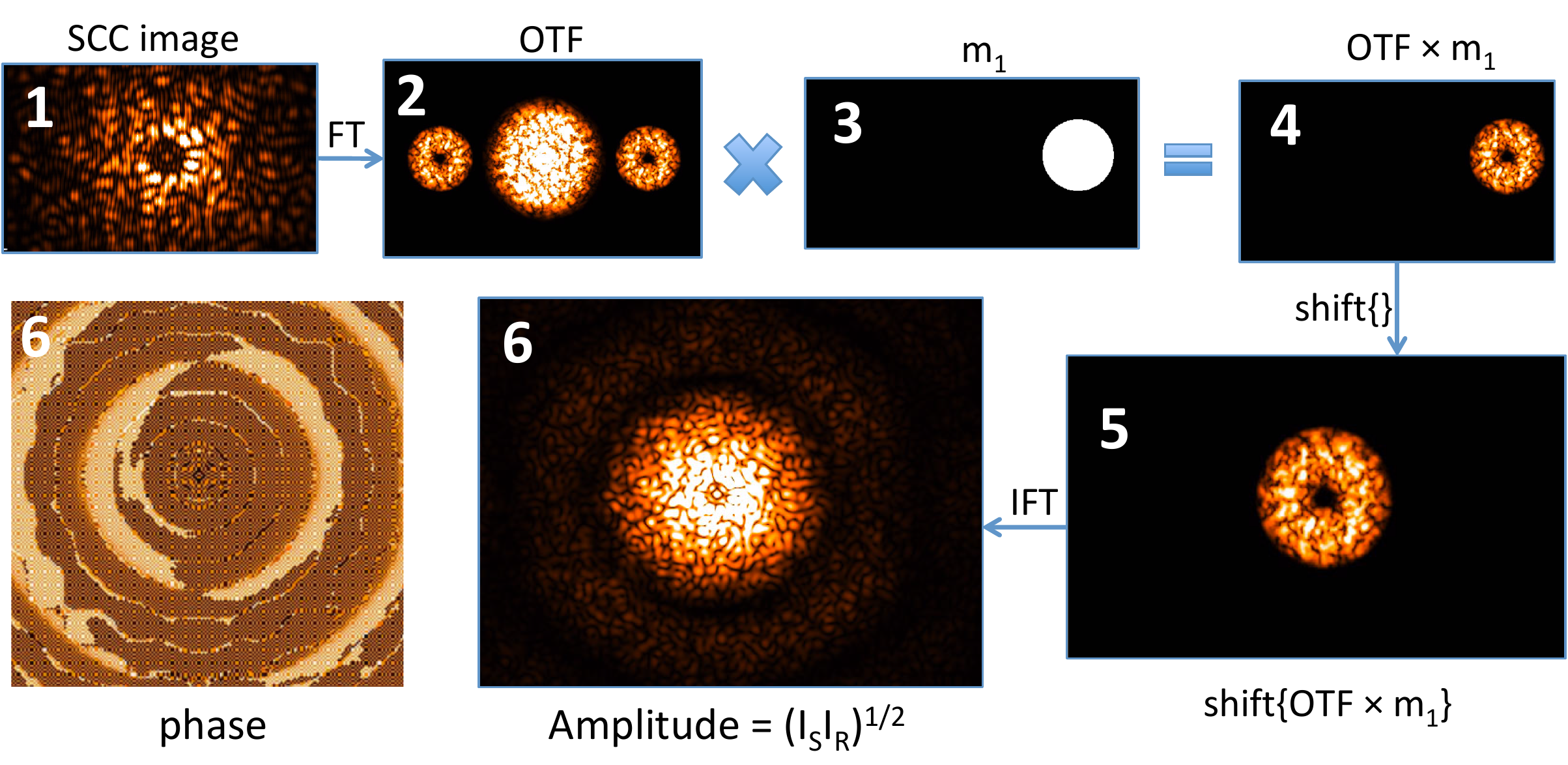}
(a)
\end{center}
\end{minipage}
\hspace*{5pt}
\begin{minipage}[b]{0.45\textwidth}
\begin{center}
\includegraphics[width=1.0\textwidth]{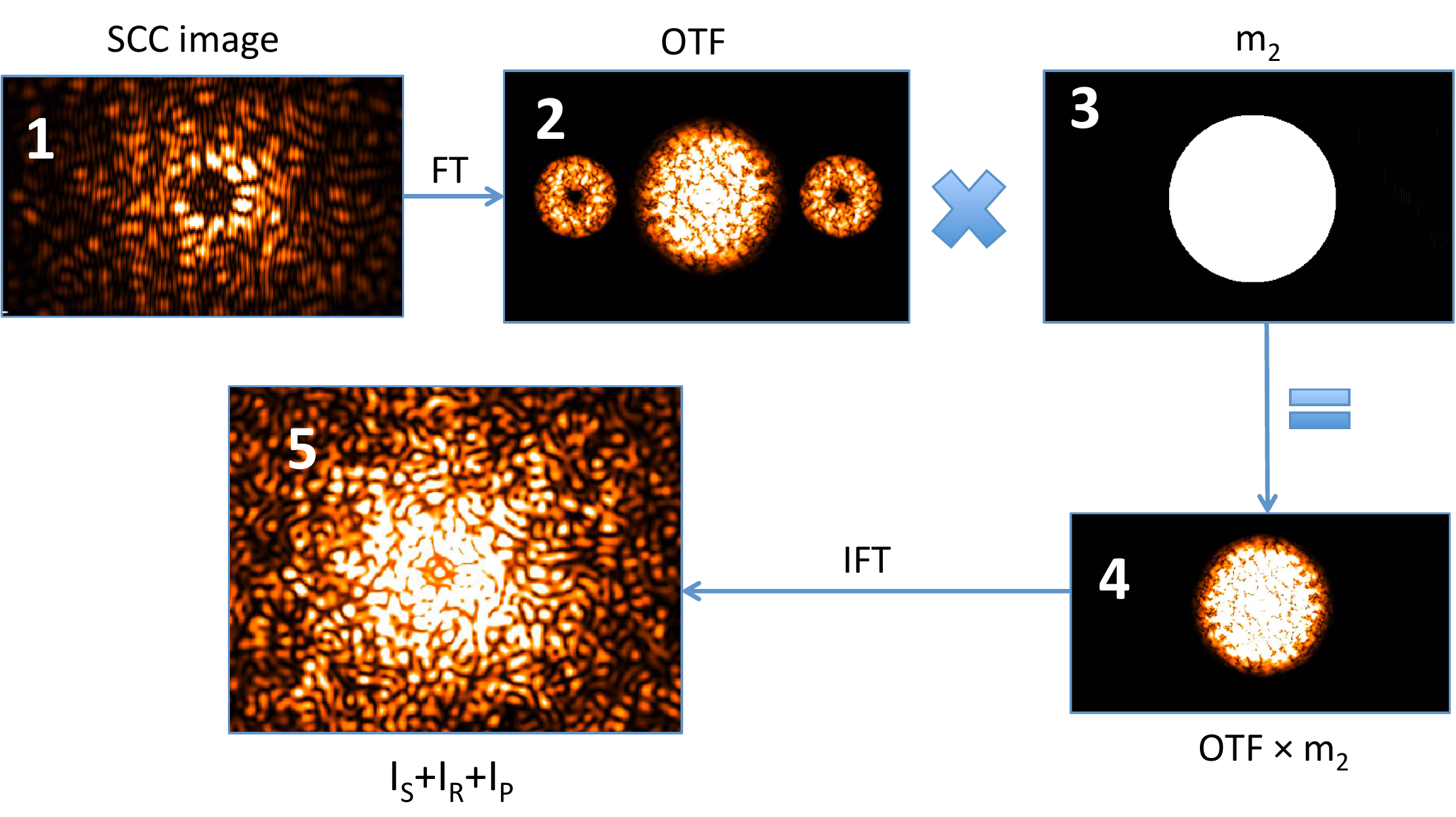}
(b)
\end{center}
\end{minipage}
\caption{An illustration, adapted from G1, of the standard SCC wavefront sensing (a) and filtering (b) algorithms. Each algorithm uses the coronagraphic SCC image in step 1 to generate a Fourier-filtered version of this image. In panel a, step 6 produces the complex-valued $I_-$. This process in panel a of converting step 1 to step 6 is represented in equation \ref{eq: fringe_snr} by the operator ``$F_{I_-}\{\}$.'' For both panels, ``FT'' and ``IFT'' represent the Fourier transform and inverse Fourier transform, respectively. In panel a, ``m$_1$'' is a binary mask to isolate the complex-valued optical transfer function (OTF; only the amplitude is shown in the above figure) sidelobe, and the ``shift\{\}'' operator repositions the isolated OTF sidelobe to the center of the Fourier plane. In panel b, ``m$_2$'' is a binary mask to isolate the central lobe of the OTF, removing the fringes but keeping the pinhole PSF and exoplanet terms.}
\label{fig: I_minus}
\end{figure}

\begin{figure}[!h]
\centering
\includegraphics[width=0.6\textwidth]{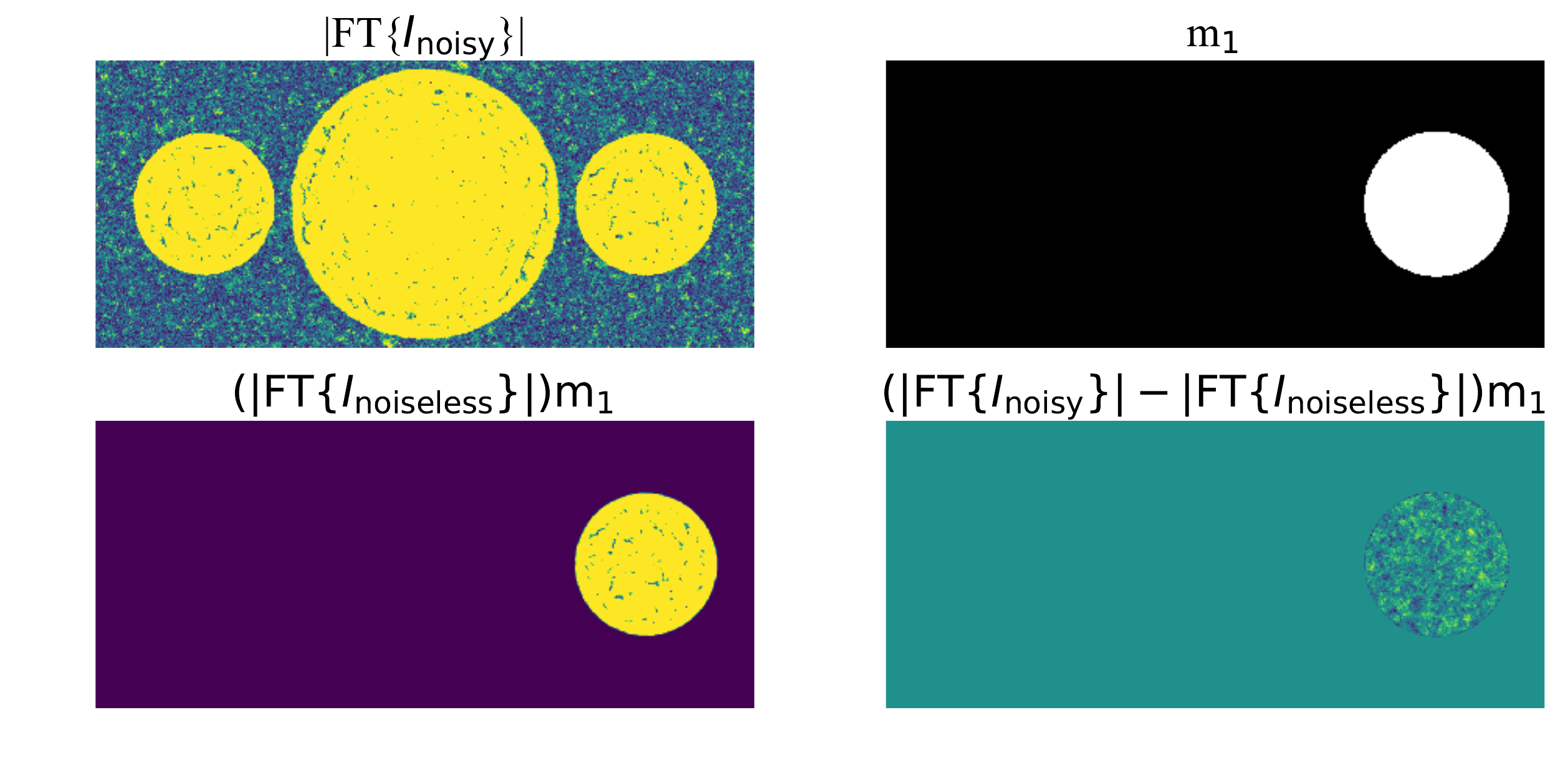}
\caption{An illustration of the OTF plane components of Equation \ref{eq: fringe_snr}, using the TG FPM for an input static phase screen with 25 nm rms in phase and 1\% rms in intensity. Upper left: the OTF amplitude, or modulation transfer function (MTF), from an SCC image with simulated photon noise. Upper right: an algorithmic binary mask to isolate the spatial frequencies of the fringes (as in Fig. \ref{fig: I_minus} a). Lower left: the noiseless (i.e., signal) component---at the isolated spatial frequencies of the fringes---of the upper left image, representing the numerator in equation \ref{eq: fringe_snr}. Lower right: the noise component---again at the isolated spatial frequencies of the fringes---of the upper left image, representing the denominator in equation \ref{eq: fringe_snr}.}
\label{fig: fringe_snr}
\end{figure}

Figure \ref{fig: I_minus}, adapted from G1, shows a summary of the Fourier filtering algorithms used to generate both $I_-$, which we will represent here by the operator $F_{I_-}\{\}$, and the ``un-fringed'' SCC image.
With these SCC Fourier filtering algorithms in mind, ``fringe S/N'' is the ratio between the signal and noise components of the SCC fringes (i.e., considering only the spatial frequencies isolated by m$_1$ in Fig. \ref{fig: I_minus} a). Accordingly, we defined the y-axis of Figure \ref{fig: vortex_wfe} c and d as
\begin{equation}
\text{fringe S/N}\equiv\frac{|F_{I_-}\{I_\text{noiseless}\}|}{\sigma\{|F_{I_-}\{I_\text{noisy}\}| - |F_{I_-}\{I_\text{noiseless}\}|\}},
\label{eq: fringe_snr}
\end{equation}
\begin{sloppypar}
\noindent where $I_\text{noisy}$ and $I_\text{noiseless}$ are SCC images from equation \ref{eq: scc_im} simulated with and without photon noise, respectively, and $\sigma\{\}$ is a numerical standard deviation operator. The Fourier plane equivalent of Equation \ref{eq: fringe_snr} is illustrated in Fig. \ref{fig: fringe_snr}.
Note that because of the absolute value signs, ${\sigma\{|F_{I_-}\{I_\text{noisy}\}| - |F_{I_-}\{I_\text{noiseless}\}|\}\neq\sigma\{|F_{I_-}\{I_\text{noisy} - I_\text{noiseless}\}|\}}$. Relatedly, computing ${I_\text{noisy} - I_\text{noiseless}}$ is less physically meaningful in this context; many pixels can detect zero photons in individual 1 millisecond frames, thus  producing $-I_\text{noiseless}$ as the ``noise'' component in these cases, which is less physically representative of photon noise in this ``quantum regime'' than the expression in the denominator of equation \ref{eq: fringe_snr} (thus motivating our choice for a 10 ms instead of 1 ms exposure in \S\ref{sec: contrast}). 
\end{sloppypar}
Similar to the intensity curves in Fig. \ref{fig: vortex_wfe}a and subsequent contrast curves, the numerator and denominator of equation \ref{eq: fringe_snr} are calculated as the median and standard deviation, respectively, in three pixel wide annuli as a function of image plane separation. Also note that equation \ref{eq: fringe_snr} only considers the S/N of the fringe amplitude but not the fringe phase (i.e., for a single fringe on a single speckle, the detectability of its intensity, relative to photon noise, but not its relative position on the speckle, respectively); as illustrated in the lower left panel of  Fig. \ref{fig: I_minus}, numerical phase wrapping prevents an analogous S/N analysis for fringe phase. However, there is no reason to believe that this ``fringe phase S/N'' would draw any conclusions that deviate from the results in Fig. \ref{fig: vortex_wfe} c and d, as the fringes still need to be detected above the photon noise in order to measure their relative position.

Next, as in derived in G1 (Sec. A.3), the ``photon noise limit,'' or photon noise limited-contrast, is determined by the combined effects of photon noise propagation through the two Fourier filtering algorithms in Fig. \ref{fig: I_minus}. With this context in mind and utilizing the definitions in Equations \ref{eq: scc_im} and \ref{eq: fringe_snr} and in Fig. \ref{fig: I_minus}, the y-axis of Fig. \ref{fig: vortex_wfe} e is given by
\begin{equation}
\text{photon noise limit}=5\; \sigma \left\{ \left( \text{IFT} \left\{ \text{FT} \left\{ I_\text{noisy} - I_\text{noiseless} \right\} m_2 \right\} \right) + \left( \frac{|F_{I_-}\{I_\text{noisy}\}|^2 - |F_{I_-}\{I_\text{noiseless}\}|^2}{I_R} \right) \right\}
\label{eq: phlim}
\end{equation}
\begin{sloppypar}
\noindent Relatedly, the images labeled as ``output photon noise limit'' in Fig. \ref{fig: phlim_illustration} are given by ${{\left( \text{IFT} \left\{ \text{FT} \left\{ I_\text{noisy} - I_\text{noiseless} \right\} m_2 \right\} \right) + \left( \frac{|F_{I_-}\{I_\text{noisy}\}|^2 - |F_{I_-}\{I_\text{noiseless}\}|^2}{I_R} \right)}}$. Note that $I_R$ in the above equation is an assumed simultaneous noiseless measurement of the pinhole PSF. As described in G1(Sec. A.3), although in reality the simultaneous measurement of an on-sky pinhole PSF will be noisy, this will only increase the noise contribution from the second term in equation \ref{eq: phlim}, therefore rendering equation \ref{eq: phlim} as a lower limit.
\end{sloppypar}

Lastly, related to the photon noise calculations in Fig. \ref{fig: vortex_wfe}, panels d and e and supporting text utilize the term ``fringe ratio'' and ``modified fringe ratio'' which we define as
\begin{align}
\label{eq: fringe_ratio} \text{fringe ratio}&\equiv I_R/I_S \text{, and} \\
\text{modified fringe ratio}&\equiv \tilde{I_R}/I_S, \nonumber
\end{align}
where $\tilde{I_R}$ in equation \ref{eq: fringe_ratio} is produced by numerically adjusting the intensity in the off-axis pinhole of the SCC Lyot stop plane by a piston ``fudge factor'' (without changing the wavefront in the central Lyot stop pupil) between values both smaller and larger than the natural, unadjusted fringe ratio value.
\bibliography{refs}

\end{document}